\documentclass[sigconf]{acmart}

\usepackage{booktabs} 
\usepackage{color}

\setcitestyle{square}

\begin{document}
\title{Solving Poisson's Equation on the Microsoft HoloLens}

\author{Anders Logg}
\affiliation{%
  \institution{Chalmers University of Technology}
  }

\author{Carl Lundholm}
\affiliation{%
  \institution{Chalmers University of Technology}
  }

\author{Magne Nordaas}
\affiliation{%
  \institution{Chalmers University of Technology}
  }


\begin{abstract}
We present a mixed reality application (HoloFEM) for the Microsoft HoloLens. The application lets a user define and solve a physical problem governed by Poisson's equation with the surrounding real world geometry as input data. Holograms are used to visualise both the problem and the solution. The finite element method is used to solve Poisson's equation. Solving and visualising partial differential equations in mixed reality could have potential usage in areas such as building planning and safety engineering. 
\end{abstract}

%
%

\ccsdesc[500]{Human-centered computing~Mixed / augmented reality}
\ccsdesc[300]{Applied computing~Mathematics and statistics}
\ccsdesc[100]{Mathematics of computing~Mathematical software}
\ccsdesc[100]{Mathematics of computing~Partial differential equations}
 
\keywords{HoloLens, Poisson's equation, finite element method, FEniCS}

\copyrightyear{2017} 
\acmYear{2017} 
\setcopyright{rightsretained} 
\acmConference{VRST '17}{November 8--10, 2017}{Gothenburg, Sweden}\acmDOI{10.1145/3139131.3141777}
\acmISBN{978-1-4503-5548-3/17/11}

\maketitle

\section{Introduction}

We develop an application, called HoloFEM, for solving Poisson's equation with the finite element method (FEM) using Microsoft's mixed reality glasses HoloLens \cite{hololens}. 
The aim is to set up and solve a partial differential equation (PDE) in the real world geometry surrounding the HoloLens user, and then visualise the computed solution on top of the real surroundings in mixed reality.

Partial differential equations are used to model many physical processes, such as fluid flow, heat transport and electromagnetic fields to name a few. 
We consider here the Poisson equation, which serves a prototypical example of a PDE. This equation models steady-state diffusion and electrostatic potential.
The Poisson problem is mathematically formulated as: Find the solution $u: \Omega \to \mathbb{R}$ such that  
\begin{equation}
\left\{
\begin{split}
- \Delta u & = f && \mbox{in } \Omega, \\
u & = g && \mbox{on } \partial\Omega,
\end{split} 
\right. \label{eqpoisson}
\end{equation}
where $\Omega \subset \mathbb{R}^3$ is the solution domain, $\Delta$ is the Laplace operator, $f$ is a given source, $g$ is a given boundary value of the solution, and $\partial\Omega$ is the boundary of $\Omega$. 

Since various PDEs often show up in modelling and engineering problems, there could be a potential use for mixed reality PDE-solving software that allows a user to define a problem, compute a solution, and study it on the spot. Say for example that we would like to know how a dangerous substance spreads in a room after a leak has sprung. The heat equation could be used as a simplistic model for describing this situation, potentially making applications like HoloFEM useful in building planning and safety engineering. 

Augmented, mixed, and virtual reality are already used in engineering, architecture, and design, see for example \cite{Bendinger:2004aa, Heuveline:2011aa}, but to our knowledge there is currently no other mixed reality PDE-solving software.

\section{Technical description}

The workflow of the application HoloFEM has three main stages.
The first one is the meshing stage, where a computational mesh is generated from the environment. 
This is followed by the simulation stage, where the mathematical problem is formulated and solved.
Finally, the solution is visualised in the last stage.

\subsection{Meshing}
The Microsoft HoloLens can scan the user's surroundings and extract a discrete representation of the geometry, in the form of a surface mesh.
This mesh is not adequate for numerical computations -- firstly, it is a surface mesh, while we need a volume mesh to solve \eqref{eqpoisson}, and secondly, the mesh quality is rather poor.

\begin{figure}[h]
\includegraphics[scale=0.048]{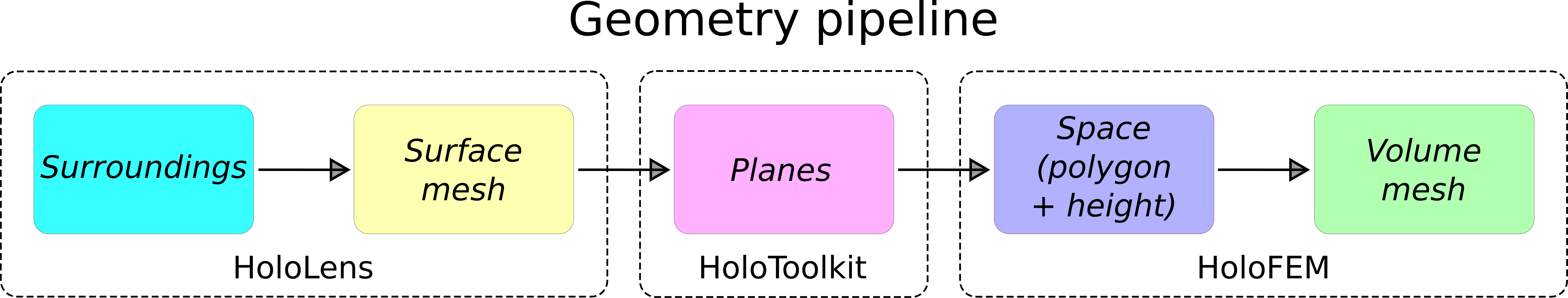}
\caption{The geometry pipeline shows the steps in going from a spatial scan of the surroundings to a volume mesh.}
\label{figgp}
\end{figure}

\begin{figure}[H]
\includegraphics[scale=0.05]{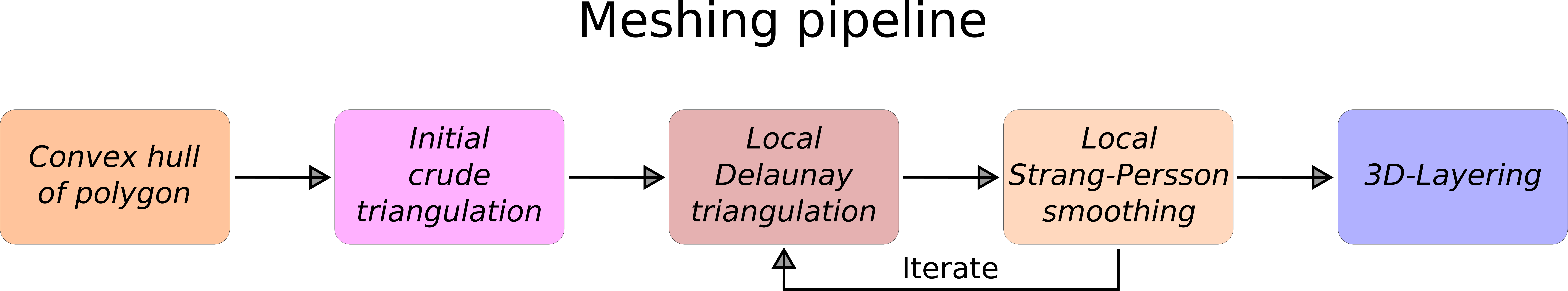}
\caption{The meshing pipeline demonstrates the generation of a volume mesh from a space, i.e., the last arrow in the geometry pipeline.}
\label{figmp}
\end{figure}

Instead, the main planes are extracted from the surface mesh. These will represent the walls, the floor and the ceiling of the room in which the user is located. 
From this geometric representation, the computational volume mesh is constructed. 
The procedure so far is summarised in Figure \ref{figgp}. The steps in the mesh generation are shown in Figure \ref{figmp}. 

\subsection{Simulation}
The user may configure the problem parameters by placing point sources in the surroundings and setting boundary conditions on the walls, floor, and ceiling of the room. 
This is done in a similar fashion to how holograms are usually placed with the HoloLens. 

\begin{figure}[H]
\includegraphics[scale=0.05]{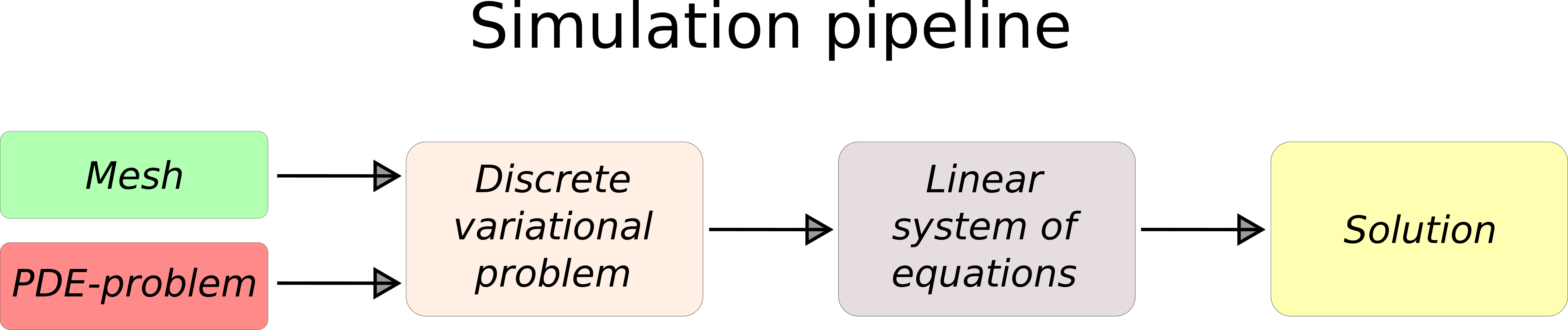}
\caption{The simulation pipeline outlines how a mesh and PDE-problem are used to obtain a solution with FEM.}
\label{figsp}
\end{figure}

When the user is satisfied with the problem specification, the problem is discretised with the finite element method. 
The FEniCS form compiler \cite{ffc,logg2012automated} is used to generate code for the finite element assembly, see Figure \ref{figsp}. This means that this part of the program can easily be generalised to handle other PDEs.
The assembled sparse linear system can be solved with standard techniques, e.g., preconditioned Krylov subspace methods.

\subsection{Visualisation}
The visualisation of the numerical solution takes advantage of the mixed reality technology of the HoloLens.
Holograms are placed at nodal points of the computational mesh, or at the cell centers, and are superimposed on the real world background.
This simplistic approach suffices for the current prototype.
More sophisticated approaches could be developed specific to the engineering applications considered.

\section{Demo overview}

\begin{figure}[H]
\includegraphics[scale=0.084]{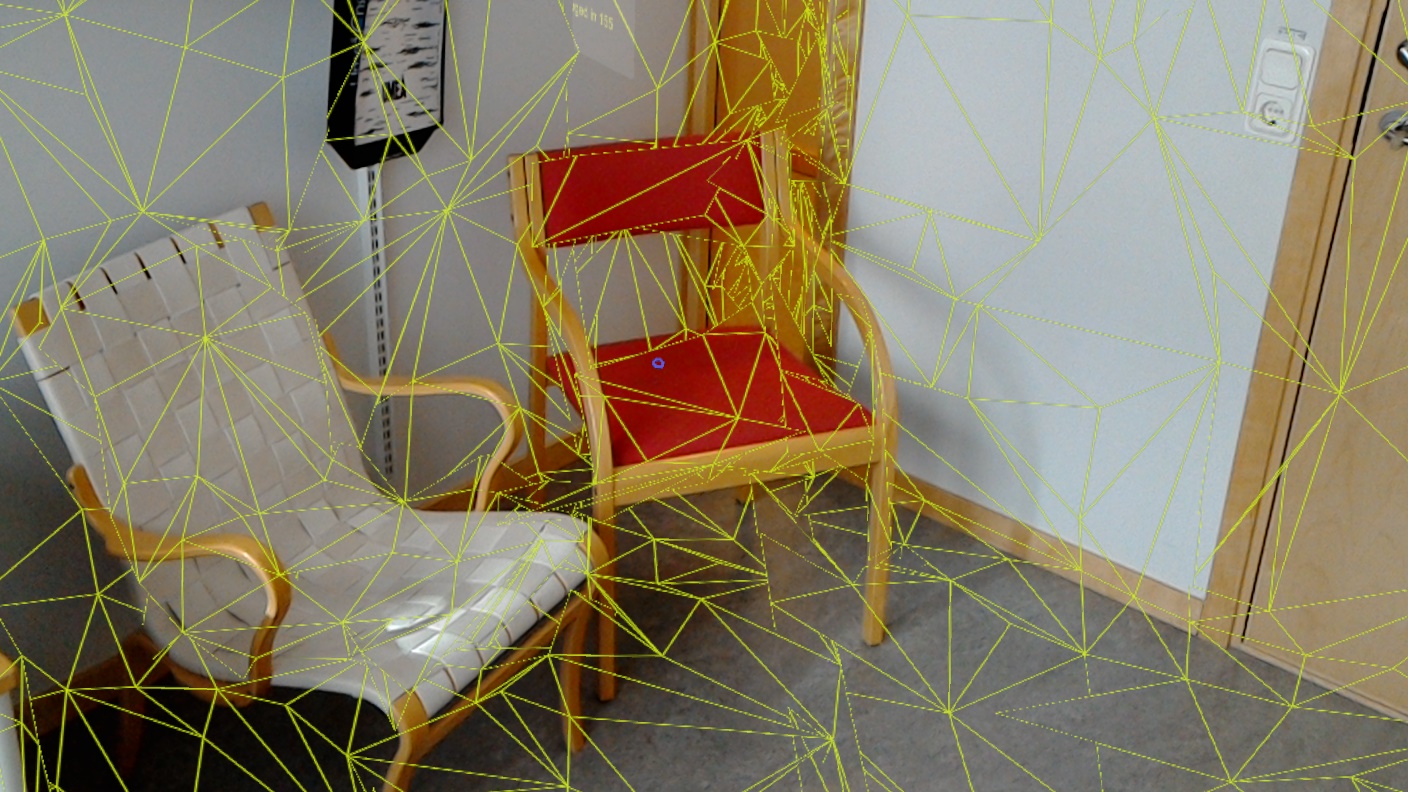}
\includegraphics[scale=0.084]{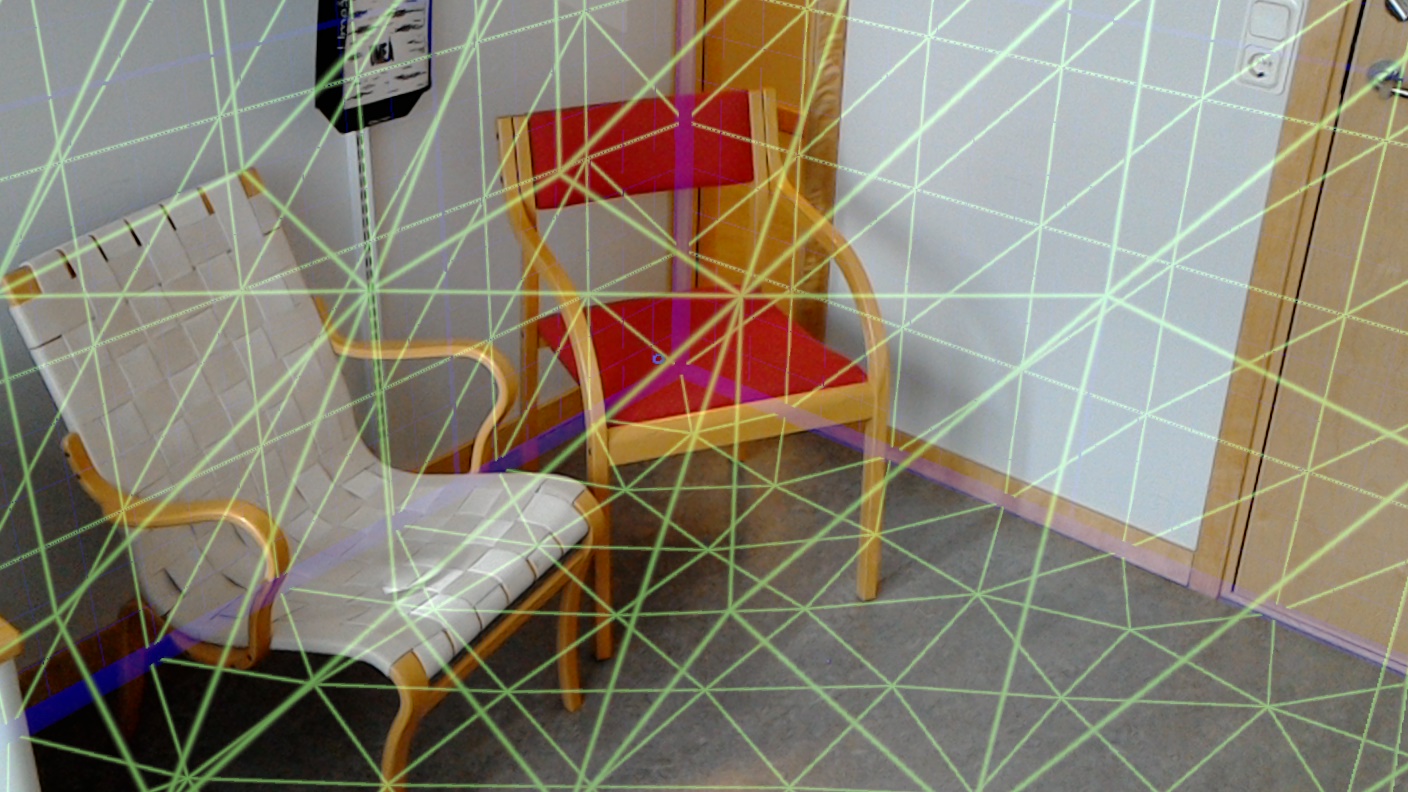}
\caption{Meshes. \emph{Left}: Surface mesh (yellow) of the surroundings. \emph{Right}: Space (blue) and generated volume mesh (green).}
\label{figs2m}
\end{figure}

In its current state the application HoloFEM works mainly with voice commands. A user, wearing the HoloLens, starts by saying ``Scan room''. This initiates the scanning phase in which the surroundings are scanned by looking around the room. During the scanning phase a surface mesh is produced that shows what surfaces have already been scanned, see the left part of Figure \ref{figs2m}. After the scanning phase is completed, the approximative geometric representation of the room (a prism with polygonal base) is automatically created. A tetrahedral mesh is then generated with the voice command ``Generate mesh''. In the right part of Figure \ref{figs2m}, parts of a space and a volume mesh are displayed.

\begin{figure}[H]
\includegraphics[scale=0.084]{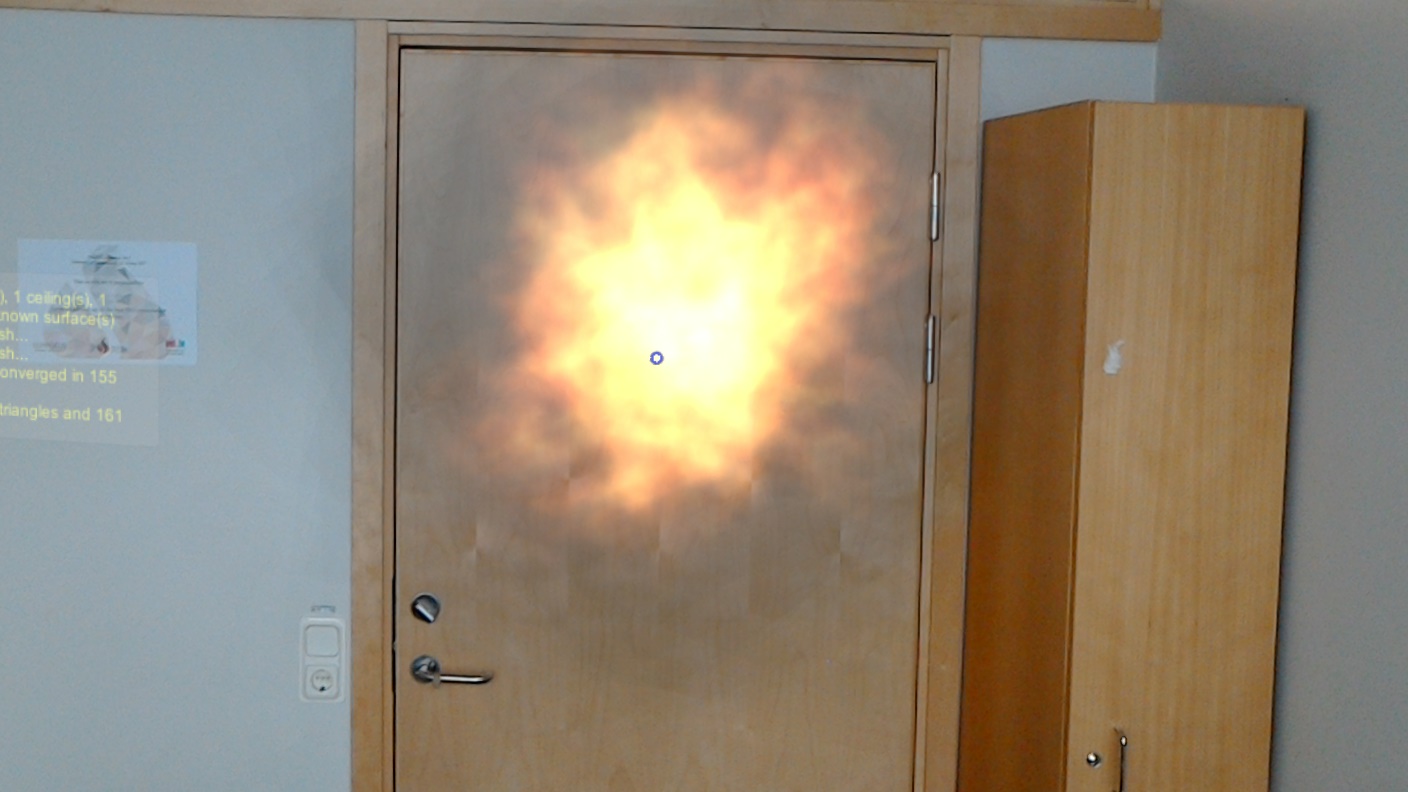}
\includegraphics[scale=0.084]{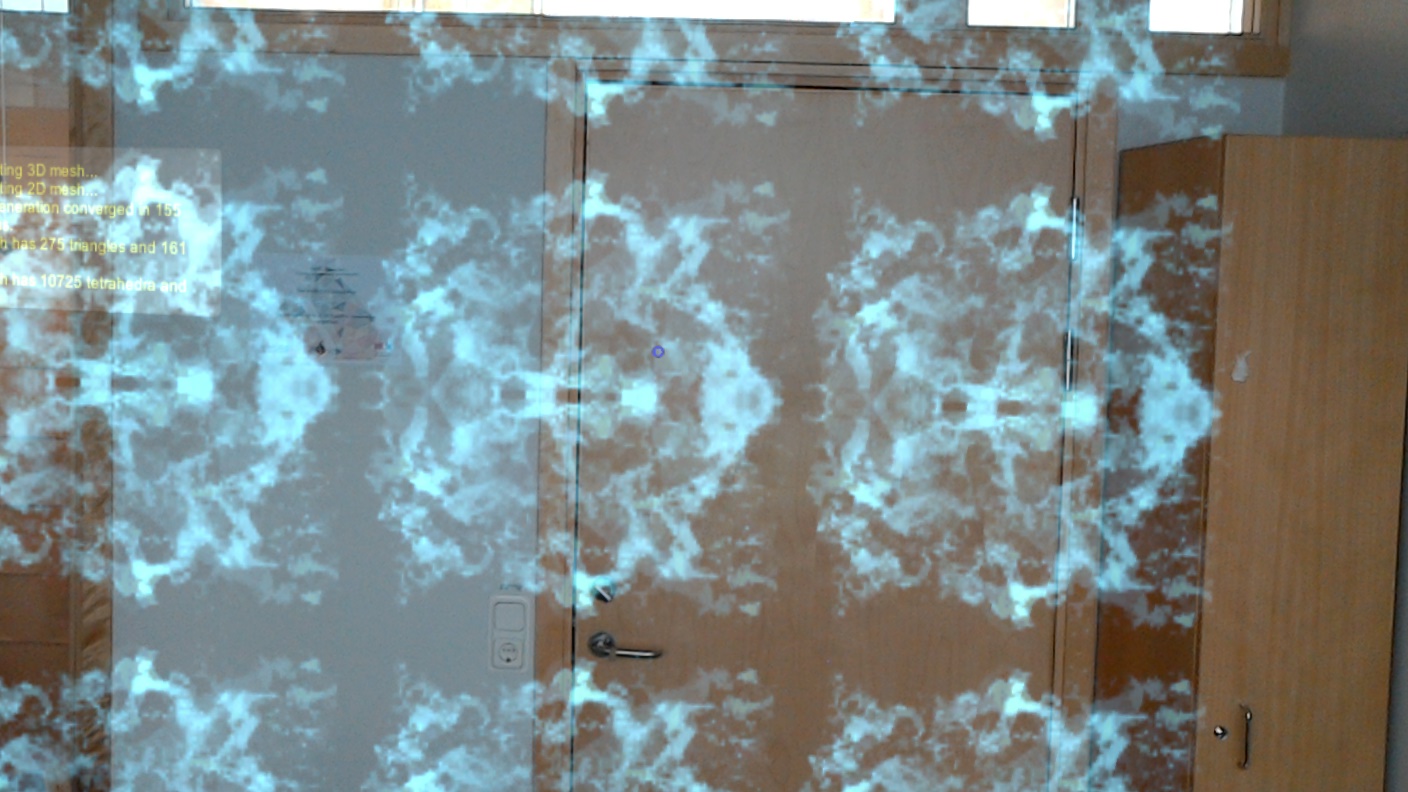}
\caption{Problem data represented by holograms. \emph{Left}: Source (fire ball) placed in room. \emph{Right}: Zero value boundary conditions (ice patches) on wall.}
\label{figproblemdata}
\end{figure}

When the mesh has been generated, the user may define additional problem data. The voice command for defining a source is ``Create source''. This places a source in front of the user. The voice command for defining boundary conditions is ``Set boundary value''. This sets the solution to be zero on the wall the user is looking at. In Figure \ref{figproblemdata}, visualised problem data are shown.

\begin{figure}[H]
\includegraphics[scale=0.084]{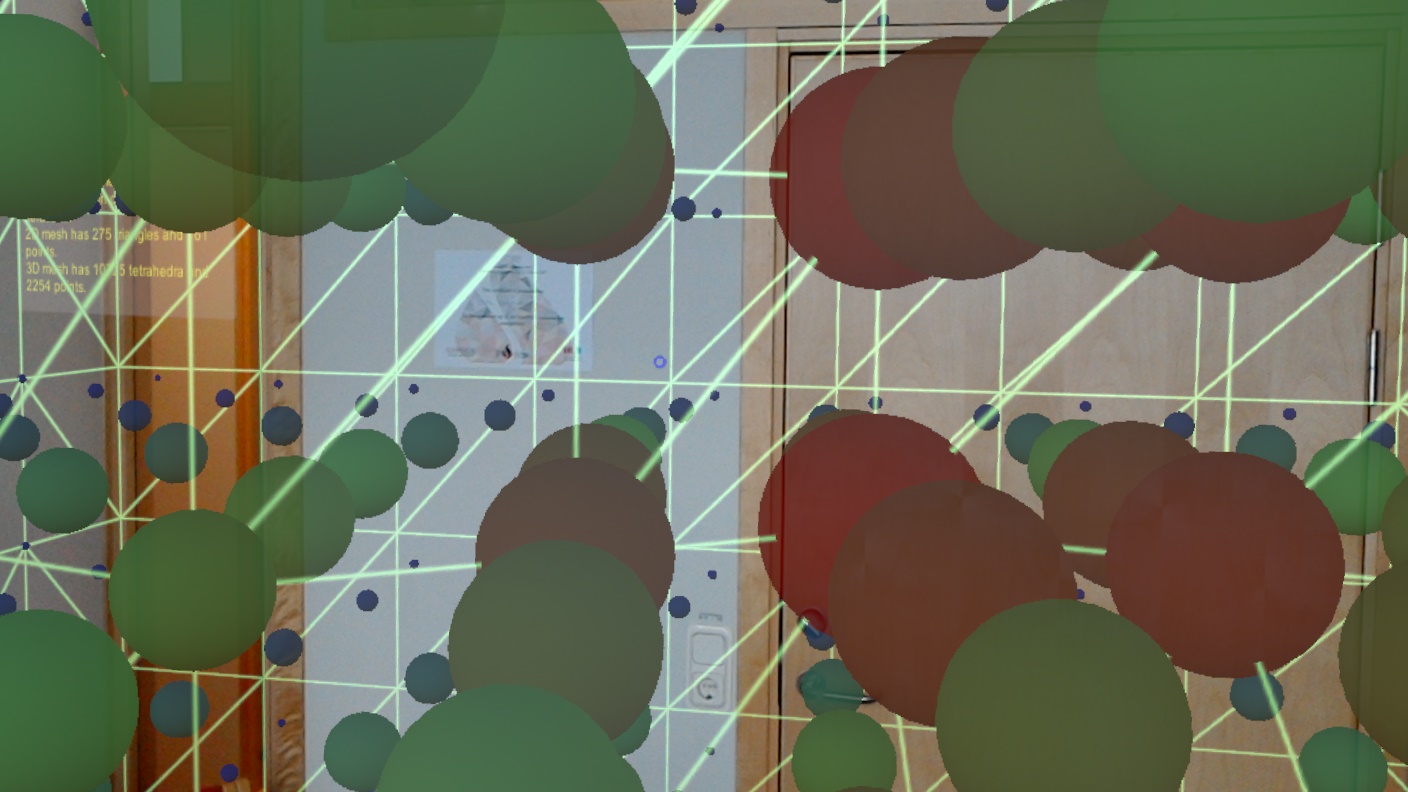}
\includegraphics[scale=0.084]{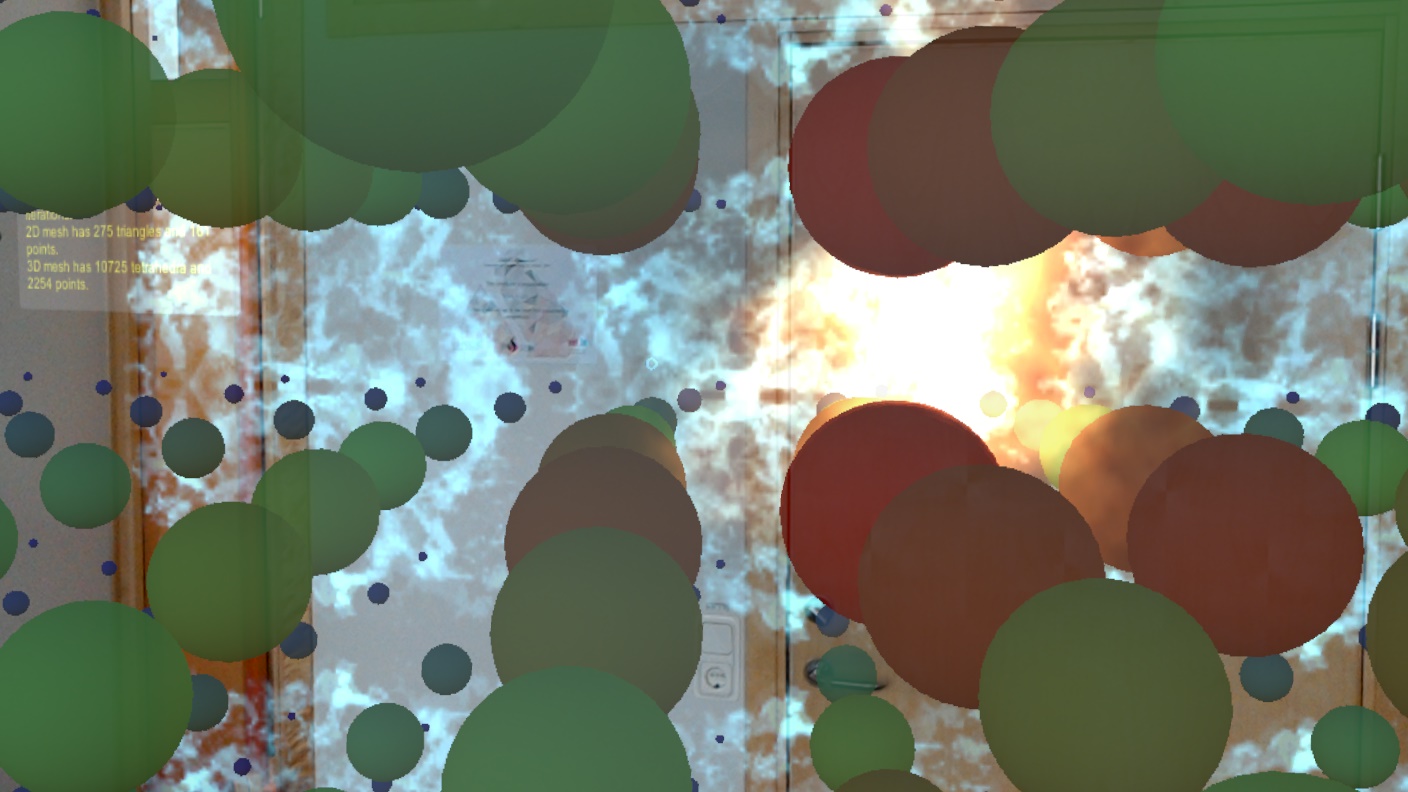}
\caption{Solution represented by spheres. \emph{Left}: Solution and tetrahedral mesh used in simulation. \emph{Right}: Solution together with problem data used in simulation.}
\label{figsolution}
\end{figure}

Once the desired problem data have been defined it is time to solve the problem. The voice command for that is ``Solve problem''. After the problem has been solved, the solution is automatically visualised. The solution values at the nodal points of the volume mesh are represented by spheres. The size of a sphere is proportional to the solution value. The spheres are also coloured according to an RGB-scale, where blue represents low values and red high values. See Figure \ref{figsolution} for a visualised solution.

\section{Conclusions}
We have developed an application for solving and visualising a PDE model with the Microsoft HoloLens. 
A user wearing the HoloLens can scan the surroundings, define a mathematical model and see the numerical solution superimposed on the real world,
all within a matter of seconds. This has potential applications in building planning and safety engineering.
Future development includes extension to other PDE models and more sophisticated visualisation.

\bibliographystyle{ACM-Reference-Format}
\bibliography{holofem} 

\end{document}